\documentclass{ws-ijmpd}

\begin{document}

\markboth{G. E. Volovik}
{On de Sitter Radiation via Quantum Tunneling}

%
\catchline{}{}{}{}{}
%

\title{On de Sitter radiation via quantum tunneling }

\author{G.E. Volovik }

\address{Helsinki University of Technology,
Low Temperature Laboratory,\\ P.O. Box 5100, FIN--02015 TKK,  Finland, \\
Landau Institute for Theoretical Physics, 119334 Moscow, Russia\\
volovik@boojum.hut.fi}

\maketitle

\begin{history}
\received{Day Month Year}
\revised{Day Month Year}
\comby{Managing Editor}
\end{history}


 \begin{abstract}
 We discuss why the tunneling picture does not necessarily lead to Hawking radiation from the de Sitter horizon. The experience with the condensed matter analogs of event horizon suggests that
the de Sitter vacuum is stable against the Hawking radiation. On the other hand the detector immersed into the de Sitter background will detect the radiation which looks as thermal with the effective temperature twice larger than the  Hawking temperature associated with the cosmological
horizon.
 
 \keywords{de Sitter spacetime, Hawking radiation, Unruh effect, quantum tunneling }
\end{abstract}

 \section{Introduction}

Hawking effect was originally invented as the unusual thermal
radiation of the black hole \cite{Hawking1974}. Later it became clear that the phenomenon
of Hawking radiation from the event horizon belongs to a broad class ofÊÊ quantum
phenomena, related to the behavior of the quantum vacuum in the nontrivial
space-time. Radiation by an object  moving along some trajectory in the quantum vacuum also belongs to this class.  This includes as  particular cases  the Unruh effect for hyperbolic motion \cite{Unruh1976}; radiation by an object moving along the circular trajectory, or by a dielectric cylinder rotating in the quantum vacuum first suggested by Zeldovich \cite{zeld}. The objects moving  in the quantum vacuum also serve as Unruh-De Witt detectors of the quantum correlations of the vacuum fields along the trajectory. Based on the superradiance phenomenon experienced by a body rotating in the quantum vacuum, Zeldovich predicted (even before Hawking) that  a Kerr black hole  will show both amplification
and spontaneous emission \cite{zeld,staro}. A rotating black hole provides an example of the ergoregion -- the region in which particles have negative energy; the behavior of the quantum vacuum in the presence of an ergoregion is another interesting problem of that kind. 

All these quantum effects may take place not only
in the atsrophysical objects: for example sound waves propagating in a
moving medium are similar to the light propagating in the curved space \cite{Unruh1981}. It is
possible to create artificial horizons and other nontrivial spacetimes including those which cannot occur in astrophysics, for example the white hole horizon, and also to study radiation by objects moving in the quantum `vacuum' of a superfluid liquid or  Bose-Eistein condensate (BEC). 

The problem which is relevant for our expanding Universe is related to the cosmological horizon. Does it radiate?  If yes,  how this radiation influences the expansion rate. 
 In particular, there is a problem with the de Sitter space-time, since the decay via Hawking radiation (if it occurs)  should lead to the relaxation of the cosmological constant.
The  stability problem of the quantum vacuum in the de Sitter background is discussed in recent papers
\cite{Polyakov2008,Higuchi2008}.

Here we consider the problem of Hawking radiation in the de Sitter background using 
the semiclassical tunneling approach developed for the astronomical black holes
by Parikh and Wilczek \cite{ParikhWilczek}, and for their condensed-matter analogs in Refs. 
\cite{Volovik1999a,Volovik1999b}.  In this method the tunneling exponent is calculated using the semiclassical analysis of transition between classical trajectories inside and outside the horizon in the Painlev\'e-Gullstrand coordinate frame \cite{Painleve}.   
For condensed matter analogs, the Painlev\'e-Gullstrand coordinates naturally emerge 
in the effective gravity induced by the flow of liquids, superfluids or BEC 
which allows us to simulate the black hole and cosmological horizons \cite{Unruh1981,Volovik1998,Volovik2003,Uhlmann2005}.  The semiclassical tunneling description of the (analog of) Unruh effect can be found in Chapter 6.15 of the book \cite{Volovik1992}. The method was applied to the (analog of) Hawking effect  in Refs. \cite{Volovik1999a,Volovik1999b}, and extended to the phenomena related to ergoregion and radiation by a rotating objects in Refs.  \cite{CalogeracosVolovik1999,Volovik1999b}.
Application of this method to de Sitter background can be found in Refs. \cite{Parikh2002,Shankaranarayanan2003,Medved2008,Sekiwa2008}.

Since in the analog systems the properties of the quantum vacuum are well known, one may conclude in what cases the quantum tunneling approach is justified.
Based on this experience we shall discuss the difference between the properties of the vacuum in the presence of the black hole horizon and the de Sitter cosmological horizon. At first glance the systems are similar, and both case the tunneling exponent is expressed via the corresponding Hawking temperature determined by gravity at the horizon.
In case of the cosmological horizon in the de Sitter space-time, the corresponding Hawking temperature  $T_{\rm H}=\hbar H/2\pi$, where $H$ is the Hubble parameter in the de Sitter universe.

However, the condensed matter approach differentiates between these two space-times,  suggesting  that radiation there belong to different types of phenomena related to quantum vacuum. One group deals with the real radiation from the vacuum. This radiation does not depend on whether there is a detector which measures the radiation or not. Radiation leads to the dissipation and decay of the horizon: the black hole loses mass and or angular momentum. In the analogous condensed matter systems, the dissipation via radiation leads for example to shrinkage of the region between the black hole and white hole horizons.

In the other group of phenomena, the detector  plays the crucial role: without the detector, the vacuum is not radiating and remains stable. The radiation occurs only in the presence of the detector and due to the interaction of the quantum vacuum with the detector. This group of phenomena includes
the Unruh effect, describing radiation by an object accelerating in the vacuum or rotating in the vacuum. In superfluids and atomic BEC this is radiation caused by an object moving in the superfluid `vacuum'.

We suggest that radiation in the de Sitter background belongs to second group
of phenomena. The pure de Sitter universe is not radiating,
i.e. it  is stable towards the radiation, while a detector immersed into the de Sitter vacuum  will register the radiation. We discuss  this effect using  an atom
as a detector. The ionization rate of an atom in the de Sitter background looks as thermal, with the effective activation temperature 
$T=\hbar H/\pi$.  This temperature is not related to the cosmological horizon and is twice larger than the the Hawking temperature  $T_{\rm H}=\hbar H/2\pi$.

 \section{Hawking temperature and tunneling exponent}
 \label{TunnelingExponent}

Both  the  black hole horizon and the cosmological horizon are described by 
the so-called ``fluid'' metric which is characterized by the ``velocity'' field ${\bf v}$:
\begin{equation}
ds^2=g_{\mu\nu}x^\mu x_\nu= -c^2dt^2 + \left(d {\bf r}- {\bf v}dt\right)^2~.
\label{FluidMetric1}
\end{equation}
The black hole  at the end of the gravitational collapse is described by the Painleve-Gullstrand metrics which corresponds to the radial flow field in the form:
   \begin{equation}
 {\bf v}({\bf r})=v(r)\hat{\bf r} ~~,~~v(r)= -c~\sqrt{\frac{r_g}{r}}~~,
\label{Schwarzschild}
\end{equation}
where $r_g$ is the radius of the black hole horizon.
The de Sitter spacetime is also characterized by the radial velocity field
\begin{equation}
{\bf v}({\bf r})=v(r)\hat{\bf r} ~~,~~v(r) = c \frac{r}{r_g} ~,
\label{FRWfluid}
\end{equation}
where $r_g$ is the radius of cosmological.
In both cases the particle spectrum is given by
\begin{equation}
E({\bf p},{\bf r})= \pm c p + {\bf p}\cdot {\bf v} ~,
\label{Spectrum}
\end{equation}
where the first term is the spectrum in the frame  comoving with the vacuum, while the last term plays the role of the Doppler frequency shift. In condensed matter, this fully corresponds to the spectrum of massless `relativistic' quasiparticle (say, phonon) in a moving superfluid or BEC.
Equation $E({\bf p},{\bf r})=E={\rm const}$ determines the radial trajectories of massless particles: 
\begin{equation}
p_r(r)=\frac{E}{v(r)\pm c} ~.
\label{Trajectories}
\end{equation}
The integration contour contains the path along the semicircle around the pole at $r=r_g$ where $|v(r)|=c$, which relates the tunneling probability to the imaginary part in the semiclassical action $w\propto \exp(-2{\bf Im}~S)$:
\begin{equation}
2{\bf Im}~S=2{\bf Im}\int dr ~p_r(r)=2E~ {\bf Im}\int\frac{dr}{v(r)\pm c} = \frac{2\pi E} {dv/dr|_{r=r_g}}~.
\label{Action}
\end{equation}
The latter is interpreted as the thermal  radiation from the horizon with Hawking temperature
\begin{equation}
T= \frac{\hbar}{2\pi } \frac{dv} {dr}\Big |_{r=r_g}~.
\label{HawkingT}
\end{equation}
For the de Sitter Universe with its $v(r)=Hr$, the corresponding temperature
would be
\begin{equation}
T= \frac{\hbar H}{2\pi }~.
\label{HawkingTDS}
\end{equation}

 \section{Detector in de Sitter background}
\label{DdSb}

Let us now introduce the detector. We consider the simplest  example of detector -- an atom immersed into the de Sitter background -- and estimate the ionization rate of the atom caused by the de Sitter expansion. This is analog of decay of a composite particle in the de Sitter background discussed
by Nachtmann \cite{Nachtmann1967}: a neutral atom is decaying into electron and ion.
The atom which plays the part of the detector is assumed to be at rest in the comoving reference frame. In the reference frame of the atom its position is at the origin, $r=0$.
The electron  bonded to an atom   absorbs the energy from the gravitational field of the de Sitter background, which is sufficient to escape from the electric potential barrier that originally confined it.
If the electron is originally 
sitting at the energy level $E_n$, then the  ionization potential $\epsilon_0=-E_n$. If the  ionization potential is much smaller than the electron mass,  $\epsilon_0\ll m$ , one can use the non-relativistic quantum mechanics to estimate the tunneling rate through the barrier. 
The corresponding radial trajectory $p_r(r)$ is obtained from the classical equation   $E(p_r,r)= -\epsilon_0$,
which in the non-relativistic approximation reads 
 \begin{equation}
 -\epsilon_0=\frac{p_r^2(r)}{2m} -  p_r(r)Hr~,
\label{RadialTrajDS}
\end{equation}
where as before $p_r$ is the radial momentum of electron, and the last term 
is the corresponding Doppler shift in the de Sitter Universe.  This gives the following radial trajectory of electron:
\begin{equation}
p_r(r)=mHr\pm \sqrt{m^2H^2r^2 -2m\epsilon_0}~.
\label{RadialTrajDS2}
\end{equation}
The momentum $p_r$ is imaginary in the region $0<r<r_0$, where 
\begin{equation}
r_0^2=\frac{2\epsilon_0}{mH^2}~.
\label{RegionBarrier}
\end{equation}
This  demonstrates that there is an energy barrier between the position of the electron in the atom, i.e. at $r=0$, and the position of the free electron with the same energy at $r=r_0$.
Since we assume that $\epsilon_0\ll m$, one has $r_0\ll r_g=c/H$, which means that tunneling occurs well within the horizon. We also assume that $H\ll \epsilon_0 (\epsilon_0/m)^{1/2} \alpha^{-1}$, which
allows us to neglect the region close to the origin where the contribution of the Coulomb potential $-\alpha/r$ to Eq.(\ref{RadialTrajDS}) is important.

The imaginary part of the action
\begin{equation}
2{\bf Im}~S=2{\bf Im}\int dr ~p_r(r)=2mH\int_0^{r_0}dr \sqrt{r_0^2-r^2}=\frac{\pi}{2}mHr_0^2=
\frac{\pi\epsilon_0}{H}~,
\label{IonizationExponent}
\end{equation}
gives the probability of ionization 
\begin{equation}
w\propto \exp(-2{\bf Im}~S) 
=\exp\left(-\frac{\pi\epsilon_0}{H}\right)~,
\label{IonizationProbability}
\end{equation}
This corresponds to thermal activation of an atom by a bath with effective temperature 
\begin{equation}
T_{\rm activation}=\frac{\hbar H}{\pi}~.
\label{ActivationT}
\end{equation}
This temperature is twice larger than
the corresponding Hawking temperature
in Eq.(\ref{HawkingTDS}), adding another twist to the factor
of 2 problem discussed for the Hawking and Unruh effects
(see Refs. \cite{AkhmedovaPillingGillSingleton2008,Pilling2008} and references therein).
It is important that the effective temperature (\ref{ActivationT}) has nothing to do with 
the existence of cosmological horizon, since the energy barrier 
is situated well within the horizon: $r_0\ll r_g=c/H$ if $\epsilon_0\ll m$.
That is why the possible subtleties, which may influence the semiclassical 
tunneling approach in the presence of horizon 
and restore the `correct' factor \cite{Akhmedova2-2008,Akhmedov2008},
are irrelevant here. 

This difference between  Eq.(\ref{HawkingTDS}) and Eq.(\ref{ActivationT})
simply demonstrates, that one should resolve between the effective temperatures
which characterize two different types of radiation:
(i) Radiation which leads to the decay of the vacuum; it exists even if there is no detector 
which could measure the radiation.  (ii) Radiation which is caused by the interaction of the detector with
the vacuum; it does not lead to the vacuum decay, but influences only the detector.

The class (i) contains Hawking radiation from the black hole horizon and Zeldovich-Starobinsky
radiation from the ergoregion of a rotating black hole  \cite{zeld,staro}.
In condensed matter this corresponds to radiation of phonons by an inhomogeneous supersonic flow. 

The class (ii) contains the Unruh effect considered as the radiation by an accelerated body, or 
by a body moving along the circular orbit (rotational Unruh effect). 
In condensed matter this corresponds to radiation of phonons or other quasiparticles by a particle (or small body) moving in superfluid liquids or in atomic BEC. It occurs in case of accelerated motion,
 or motion along the circular trajectory, or even if a particle (or small body) moves with constant velocity, which exceeds the Landau critical velocity. The semiclassical tunneling approach to such a radiation is discussed in Appendix.

We suggest that the de Sitter background supports only the second type of radiation.

 \section{de Sitter background vs black hole background}

The tunneling exponent for the cosmological horizon \cite{Parikh2002} and for the black-hole horizon  are given by the same equation (\ref{Action}). However, this does not imply that the radiation would  necessarily take place. The possibility of radiation essentially depends on the properties of the vacuum in these two spacetimes. To clarify that we follow the reasoning in Ref. \cite{Volovik1999a,Volovik2003} for fermionic vacua. In both cases one can define the comoving vacuum, i.e. the vacuum as seen by the local observer moving with velocity ${\bf v}({\bf r})$.
In his/her frame, the spectrum of particles is
\begin{equation}
E_{\rm comoving}({\bf p})= \pm c p  ~.
\label{Spectrum}
\end{equation}
In the comoving vacuum the states  with $E_{\rm comoving}>0$ are empty, while   the states  with $E_{\rm comoving}<0$ are occupied.

In the case of a black hole, there is a preferred reference frame in which the black hole is at rest.  In this reference frame the velocity ${\bf v}({\bf r})$ is stationary and thus the energy $E$ is well defined. 
This allows us to determine the true vacuum: this is the vacuum in which the states  with $E>0$ are empty, while   the states  with $E<0$ are occupied. In the region behind the horizon, this vacuum is very different from the comoving vacuum: some states which have  $E_{\rm comoving}<0$ and thus
 are occupied in the comoving vacuum,  have positive energy when considered in the preferred frame,
 $E>0$. This should lead to the decay of the comoving vacuum to the true vacuum, and this decay starts with Hawking radiation. The latter can be treated as tunneling from the occupied positive energy states inside the horizon to the empty states outside the horizon and is described by the imaginary action in Eq.(\ref{Action}) \cite{Volovik1999a,Volovik2003}.  This can be also interpreted in the following way: the black hole itself (or actually the singularity immersed in the space-time of a black hole) plays the role of the detector interacting with the vacuum. 
The same approach with tunneling from the occupied positive energy states across the horizon or ergosurface is applicable for the analog of Zeldovich-Starobinsky effect, where ${\bf v}({\bf r})={\bf \Omega}\times{\bf r}$. Here the preferred reference frame is the frame of a body rotating in superfluid vacuum with angular velocity ${\bf \Omega}$ or a particle moving  in superfluid vacuum
along circular trajectory  with angular velocity ${\bf \Omega}$
\cite{CalogeracosVolovik1999,Volovik2003} (see \ref{CircularMotion}).
 
 As for the de Sitter case,  there is no preferred reference frame because this universe
is invariant under the action of the group of symmetries of the de Sitter space-time $O(4; 1)$ 
 (see e.g. Refs. \cite{StarobinskyYokoyama1994}). While the black hole violates the translational symmetry of the vacuum, the de Sitter metric in Eq.(\ref{FluidMetric1}) is invariant under the modified de Sitter translations 
 \begin{equation}
 {\bf r} \rightarrow {\bf r} + {\bf x}_0 \exp(Ht) ~.
\label{dStranlation}
\end{equation}
This is the transformation to the reference frame centered at the 
comoving coordinate  ${\bf x}_0$. At small $Ht\ll 1$, this
transformation  represents  translation ${\bf r} \rightarrow {\bf r} + {\bf x}_0$ combined with Galilean transformation ${\bf v} \rightarrow {\bf v} + {\bf u}_0$, were ${\bf u}_0=H {\bf x}_0$.
The translational invariance implies that
there is only a unique comoving vacuum, in which the positive energy states are empty everywhere, i.e. both inside and outside the cosmological horizon;   while the  negative energy states are everywhere occupied. That is why the cosmological horizon in de Sitter universe does not necessarily lead to quantum  tunneling and decay of the comoving vacuum. Since there is no interaction
with the preferred reference frame,  the prefactor of the tunneling exponent vanishes. 

Condensed matter analogs of horizon give us another example, when the prefactor is identically zero.  For example, the radiation of a quasiparticle changes the superfluid current behind the horizon
and this violates the mass conservation law. However, radiation of simultaneously two quasiparticles is not forbidden. For the simultaneous tunneling of two particles -- which is called the co-tunneling -- the tunneling exponent is twice the tunneling exponent of
a single quasiparticle. This correspond to the effective Hawking temperature which  is twice smaller than the nominal Hawking temperature in Eq.(\ref{HawkingT}). This effect is just opposite to the case discussed in Sec. \ref{DdSb}, where the effective temperature is twice larger.

In the case of pure de Sitter vacuum, the prefactor of the tunneling exponent describing the Hawking radiation is zero due  to symmetry reasons.
 On the contrary, the detector immersed into the de Sitter background violates the translational invariance of the de Sitter space-time since there appears the preferred reference frame of the detector. The gravitational interaction of the detector with the  de Sitter background leads to the activation of the detector, which
looks as thermal activation characterized by effective temperature (\ref{ActivationT}).
This suggests that gravity is 
capable to ionize the atom embedded into the de Sitter space-time, but is not capable to create atoms from the pure de Sitter vacuum. 

This means that the pure de Sitter vacuum is stable
due to its symmery, but
the combined system  -- de Sitter vacuum + embedded inertial detector  -- is not stable and will radiate. 
This is the situation, which is similar to radiation by non-inertial detector in Minkowski space-time. Pure 
Minkowski vacuum is stable, but the system -- Minkowski vacuum + embedded non-inertial 
detector -- is not. In both cases an atom will be ionized, but the pure vacuum without  
atoms is stable.
In both cases it is the symmetry violation by detector, which is crucial. Non-inertial 
detector in Minkowski space-time  violates the Lorentz symmetry of the vacuum,
while the inertial detector in the de Sitter space-time  
violates the generalized translational symmetry of the vacuum. 

If the observer situating at the origin has many atoms in his/her detector, then observing the decay
of atoms, he/she would conclude that there is a thermal bath of something which excites the atoms. But finally  all the atoms in the detector will be  
dissociated, the ions and electrons will be radiated away, and one  obtains the pure de Sitter vacuum. This process can be described within the classical gravity (with constant classical cosmological constant) interacting with ordinary matter above the vacuum. That 
is why there is no need for relaxation of cosmological constant in this process.

 \section{Discussion}

Our consideration supports the earlier conclusion that  de Sitter Universe does not exhibit the Hawking radiation \cite{StarobinskyPrivate}; indeed, the massless vector fields, massless Dirac fermions and conformally coupled scalars do not produce the imaginary part in their action in the background of a pure de Sitter space-time.   As for the radiation of gravitons, which are not conformal invariant, this is still under discussion  \cite{GarrigaTanaka2007,TsamisWoodard2007}. Starobinsky in Ref. \cite{Starobinskii1979}  considered this problem for the local de Sitter space time, i.e. when the initial de Sitter space-time decays due to some other reasons. He found that gravitons are created in this situation, but their spectrum differs  from what is expected from the Hawking-Unruh approach.  The stability of  de Sitter solution towards radiation of  massive scalar particles has been recently discussed in Ref. \cite{Busch2008}. All this indicates that there is no Hawking radiation from the cosmological horizon of  a pure de Sitter space-time.
 
 If de Sitter space-time is not radiating, how does this correlate with Hawking temperature discussed for the de Sitter horizon? Again the symmetry of the de Sitter state is instrumental. Let us cite Starobinsky and Yokoyama \cite{StarobinskyYokoyama1994}:
"... due to the possibility of a static representation of a part of the de Sitter space-time surrounded by
the event horizon, the same state, if exists, may be also called a  thermal state with the
temperature $T = H/2\pi$ \cite{GibbonsHawking1977}. This fact, however, simply reflects the symmetries of the de Sitter space-time and does not mean that average values of operators in this state, 
are given by corresponding thermal values in the flat space-time with the Gibbons-Hawking
temperature".

The discussed tunneling picture of radiation across the horizon is based on the existence of at least two different reference frames in which the quantum vacuum is different. In the case of a black hole horizon, this condition is satsified. However, there is another important factor in the game: how the interaction of the reference frame with quantum fields is established. The importance of this factor is seen on example
of  the Unruh-like effect -- the radiation by objects moving in the quantum vacuum.  The quantum tunneling approach to the Unruh effect is discussed  in Appendix. An object  provides the preferred reference frame, but the radiation rate though contains the tunneling exponent  also depends on interaction between the object and quantum field, and thus vanishes in the limit of vanishing interaction.

Applying the tunneling approach to the black hole one may come to the unexpected conclusion that  the situation with the Hawking radiation is not very clear.  In this approach the second reference frame, which dictates the final structure of the vacuum, is provided by singularity in the center of the black hole. The  singularity actually  plays the same role as the object moving in the quantum vacuum.  If it is so,   radiation should  depend on the details of the interaction between the vacuum fields and the singularity. This interaction could be small or even  vanish due to some symmetry. This would mean that black hole is a grey body rather  than a black body. If this is correct, the Hawking-like radiation with the thermal spectrum  but with smaller prefactor would serve as the detector which measures the quantum noise of the physical singularity. This would also pose some doubts on the area law for the entropy of the black hole. The condensed matter simulation of the event horizon also suggests  \cite{Volovik2003b} that the entropy related to horizon might be essentially smaller than the traditional Bekenstein-Hawking entropy. 
 Note that the fact that a black hole is not a black body
but a grey body in 4D (in contrast to 2D) was known long ago. The radiation frequency
spectrum is the blackbody one multiplied by black hole absorbtion coefficients coming from the effective potential outside the horizon \cite{staro}. This effect is not related to the singularity inside black hole.
However, from the condensed matter point of view, the singularity inside black hole may have important consequences both for Hawking radiation, and for the vacuum stability within the horizon.

The quantum tunneling approach to the radiation of fermions by a rotating black hole has been used in Ref. \cite{KhriplovichKorkin2001}. There are two processes: the Hawking radiation related to the horizon, and the Zeldovich-Starobinsky radiation related to the ergoregion. The Hawking part, the radiation of those fermionic particles for which the Zeldovich condition $\omega<\Omega J$ is not satisfied, can be found in recent Ref. \cite{KernerMann2008}. In general case the tunneling exponent is not described by a simple pole (as we can see on example of Unruh like effect for the accelerated object in \ref{Unruh}   and  for the object rotating in quantum vacuum in   \ref{CircularMotion}), and the radiation is not thermal. But again, as we discussed above, the tunneling exponent may have the prefactor which depends on the interaction with the singularity and thus may be highly suppressed or even be zero.

  As for the de Sitter Universe, if the interaction with the external body or with  environment is absent,  then for the symmetry reason one may expect no radiation from the cosmological horizon. The symmetry of this solution of Einstein equations makes de Sitter Universe robust to quantum corrections (see also Ref. \cite{ColeyGibbons2008}) and thus robust to the Hawking radiation. However, when an external body (or matter) is introduced, it violates the symmetry by providing the reference frame associated with the body. This gives rise to radiation, but with the prefactor depending on the interaction between the body and the radiated fields.  In principle, this may lead to an energy exchange between the vacuum and matter fields mediated by the object (or by thermal matter) and thus to the decay of the vacuum energy and cosmological constant. The phenomenological model for such a process which may occur  in the presence of matter can be found in Ref. \cite{Klinkhamer2008}, but the microscopic physics of such processes of energy exchange is not developed. 
  
All this suggests that the pure de Sitter background (i.e. without the detector, which violates the de Sitter symmetry) is stable. At least, it is stable with respect to the decay via the Hawking radiation. The more subtle effects, leading to the
decay of the de Sitter space-time, are possible, but most probably they are not related to the Hawking radiation. For 
example, Polyakov suggested that the infrared quantum fluctuations of the metric may lead to effective screening 
of the cosmological constant $\Lambda$ \cite{Polyakov1982},
and possibly to the instability of the de Sitter space-time towards the perturbations breaking the de Sitter symmetry \cite{Polyakov2008}. 

  In condensed matter, the expanding Universe can be simulated by expanding Bose-Einstein condensates \cite{Uhlmann2005}. In this case the preferred reference frame is the laboratory frame. The interaction of the expanding `superfluid vacuum' with the laboratory frame provides the real mechanism for freezing of phonon modes after crossing the acoustic horizon and the associated amplification of quantum fluctuations.

\section{Acknowledgements}
I thank Emil Akhmedov, Frans Klinkhamer, Ralf Sch\"utzhold and Aleksei Starobinsky   for illuminating discussions,  the Russian Foundation for
Fundamental Research (grant 06-02-16002-a) and the Khalatnikov--Starobinsky
leading scientific school (grant 4899.2008.2) for support.
 
 \appendix

\section{ Pair creation by accelerated object and  Unruh effect}
\label{Unruh}

Let us consider Unruh effect in terms of quantum tunneling using Refs. \cite{Volovik1992} and \cite{CalogeracosVolovik1999}. An object is moving in the vacuum with the time dependent velocity ${\bf u}(t)$. In the frame of the moving  object the metric is given by Eq.(\ref{FluidMetric1}) with  ${\bf v}=-{\bf u}(t)$.
The energy spectrum of fermions in this frame is time dependent
\begin{equation}
E_\pm({\bf p})= \pm c p - {\bf p}\cdot {\bf u} (t)~,
\label{SpectrumAcceleration}
\end{equation}
The object interacts with the fermionic field in the vacuum, and if its velocity is not constant it should radiate the corresponding fermions. 
This object  serves as the Unruh-De Witt detector which measures the correlation of the fermionic field in the frame of the body. On the other hand the static observer detects the radiation produced by the object. 
In some cases this radiation looks as thermal, with the Unruh temperature. Let us consider one of these cases.

In Ref. \cite{Volovik1992}  the time dependent velocity of linear motion was chosen, which approaches the final velocity ${\bf  u}(t\rightarrow \infty)={\bf u}_0=u_0\hat {\bf z} $ along the following trajectory:
\begin{equation}
{\bf u}(t)={\bf u}_0~{\rm th}~\frac{at}{u_0}
\label{Acceleration}
\end{equation}
where $u_0$ is some limiting velocity, which is less than (or equal to) the speed of light $c$.
This is not the hyperbolic motion, the latter would be given by
\begin{equation}
{\bf u}(t)=\hat {\bf z} ~\frac{at}{\sqrt{1+a^2t^2/c^2}}
\label{Acceleration2}
\end{equation}
 where the parameter $a$ means  the proper acceleration of the external body.
In both cases the velocity of the object does not exceed the speed of light, $u(t)<c$, that is why there is no Cherenkov radiation. The vacuum, determined as the state in which the levels with $E({\bf p})<0$ are occupied, is the same at any instant moment. However, if the body interacts with quantum fields, the time-dependent interaction disturbs the vacuum, and the body will radiate particles from the vacuum.

The tunneling process of the particle creation from the vacuum can be described  using the adiabatic perturbation theory \cite{Volovik1992}.
The  amplitude of the creation of the pair of the fergmionic particles with equal momenta  corresponds to the particle transition from the filled negative energy level 
$E_-({\bf p}_\perp, p_z,t)$ to the positive branch with the opposite $p_z$: $E_+({\bf p}_\perp, -p_z,t)=-E_-({\bf p}_\perp, p_z,t)$. In the adiabatic approximation this amplitude is given in terms of the classical action
\begin{equation}
A_{{\bf p}}=\int dt~f(t) e^{-i(S_-({\bf p},t)-S_+({\bf p},t))/\hbar}=
\int dt~f(t) e^{-2iS_-({\bf p},t) /\hbar}~,
\label{A}
\end{equation}
where
\begin{equation}
S_+({\bf p},t)=\int^tdt_1~E_+({\bf p}_\perp,-p_z,t_1)~~,~~S_-({\bf p},t)=\int^tdt_1~E_-({\bf p}_\perp,p_z,t_1)~.
\label{Action2}
\end{equation} 
The prefactor $f(t)$ depends on the details of the interaction of the fermionic field with a body. The interaction is certainly necessary to provide the real radiation by a moving body, but in the leading approximation we are first interested in the tunneling exponent.  The main contribution to the integral comes from the  stationary point $t_0$ of the exponent: $E_-({\bf p},t_0)=0$. Since we consider the velocity, which is less than the speed of light (or below the Landau critical velocity in condensed matter), the stationary point never occurs on the real time axis. So this is the tunneling process, at which the stationary point $t_0$  is situated in the complex time plane. The transition probability is given by   
\begin{equation}
w\sim \exp \left(-4~{\bf Im}\int^{t_0}( {\bf p}\cdot {\bf u} (t)-cp)dt\right)
\label{Exponent}
\end{equation}
This probability is essentially determined by the time dependence of ${\bf u}(t)$. 

The stationary point $t_0$  found for the trajectory in Eq.(\ref{Acceleration}) is 
\begin{equation}
\frac{at_0}{u_0}=\frac{\pi}{2}i+\frac{1}{ 2}{\rm ln}\frac{cp+p_zu_0}{cp-p_zu_0}~,
\label{StPoint}
\end{equation}
This gives the following creation rate in Eq.(\ref{Exponent}):
\begin{equation}
w\sim \exp\left(-\frac{2\pi u_0E_+({\bf p},t=\infty)} {\hbar a}\right)
\label{Rate}
\end{equation}
So, in the semiclassical approach the tunneling exponent for particles radiated at zero temperature by an object moving in the vacuum  with velocity in Eq.(\ref{Acceleration}),  is proportional to the Boltzmann exponent for the Doppler shifted spectrum $E_+({\bf p},t=\infty)=cp -{\bf p}{\bf u}_0$  with the effective  temperature
\begin{equation}
T_{eff}=\frac{\hbar a} {2\pi u_0}
\label{UnruhEffT}
\end{equation}
 For  $u_0=c$ this corresponds to the Unruh temperature
\begin{equation}
T_{U}=\frac{\hbar a}  {2\pi c}~.
\label{UnruhT}
\end{equation}
The prefactor in Eq.(\ref{Rate}) depends on interaction of a moving body with the vacuum fields. That is why it is not a full-ideal black body radiation, which is not surprising since the radiation should disappear in the limit of the vanishing  interaction.  On the other hand, the moving body may serve as the Unruh-De Witt detector which measures the quantum noise of the field with which the detector interacts. Emission of  the quanta of the field by the body is actually one of the tools to detect the quantum noise. Since the spectrum of radiation is proportional to the Boltzmann exponent with Unruh temperature, this allows us to say that the detector views the vacuum as thermal bath with Unruh temperature. The detector itself does not behave as a black body: it is grey body with low radiance which is determined by interaction with field.

For the trajectory in Eq.(\ref{Acceleration2}) the main contribution comes from the singularity in complex $t=t'+it''$ plane,   which is nearest to the real axis. The singularity occurs at 
\begin{equation}
t''=\frac{c}{a}  ~.
\label{StPoint2}
\end{equation}
This gives the following creation rate in Eq.(\ref{Exponent}):
\begin{equation}
w\sim \exp\left(-\frac{2\pi c^2p} {\hbar a}\right)
\label{Rate2}
\end{equation}
which corresponds to the Unruh temperature in Eq.(\ref{UnruhT}).  This suggests that the Rindler horizon is viewed by detector as the black body radiator.

\section{Circular motion}
\label{CircularMotion}

For a circular trajectory with the centripetal acceleration  $a$   one obtains the exponent  which comes from the nontrivial pole emerging in the ultrarelativistic limit \cite{Letaw1981}:
\begin{equation}
w\sim  \exp\left(-\frac{\sqrt{12}Ec}{\hbar a}\right)~~,~~ v\rightarrow c~.
\label{UltraRel}
\end{equation}
  The experimental realization of the circular  Unruh effect is the so-called Sokolov-Ternov effect \cite{SokolovTernov1963}  --  the radiative self-polarization of relativistic electrons  circulating in storage rings.   Here the spin of the electron serves as a moving detector. However, under conditions of the experiment, the tunneling exponent is not resolved since the energy
splitting of the detector is too small: $\Delta E \ll  \hbar c/a$  (for clarification see Ref. \cite{AkhmedovSingleton2007}). 
  
For the description of radiation by a rotating object,   the semiclassical  tunneling picture  is also applicable (see Chapter 31.4 in Ref.  \cite{Volovik2003} and Ref. \cite{CalogeracosVolovik1999}). Radiation takes  place  when the Zeldovich condition 
  \begin{equation}
cp< \hbar\Omega l~,
\label{ZeldovichCondition}
\end{equation}
 is satisfied,  where $\Omega$ is the angular velocity of the object and $l$ is the azimuthal quantum number of emitted photon. In the rotating frame the body is static and thus the tunneling exponent is given as in Sec. \ref{TunnelingExponent}  by the imaginary part of the action, ${\bf Im}\int dr ~p_r(r)$. But now instead of the pole one has the conventional tunneling through the classically forbidden region.
 In the rotating frame the energy is given by Eq. (\ref{Spectrum}) with $ {\bf v}=-{\bf \Omega}\times {\bf r}$.
 Taking into account that the angular momentum is quantized, and thus the azimuthal component of the momentum is $p_\phi=l/r$, one obtains the spectrum
  \begin{equation}
E(p_z,p_r,l)= \pm c p + {\bf p}\cdot {\bf v} = \pm c\sqrt{p_z^2+p_r^2+\frac{\hbar^2l^2}{r^2}} -\hbar\Omega l~,
\label{SpectrumRotating}
\end{equation}
which gives  trajectories $p_r(r)$ for fixed $E$, $p_z$ and $l$:
  \begin{equation}
p_r^2(r)= \frac{(E-\hbar \Omega l)^2}{c^2}- p_z^2- \frac{\hbar^2l^2}{r^2} ~.
\label{TrajectoryRotating}
\end{equation}
Photons with $E<0$ and thus with $cp<\hbar\Omega l$ are spontaneously radiated by the body. The created photon must tunnel from the body, which is at the position $r=R$, to the classically allowed region.
  Under condition (\ref{ZeldovichCondition}) one obtains the following tunneling exponent:
  \begin{eqnarray}
w\sim \exp\left( -\kappa l\right)~~,~~\kappa= \ln\frac{1+u}{1-u}-2u~,
\nonumber
\\
u=\sqrt{1-\left(\frac{p_rR}{\hbar l}\right)^2}=\sqrt{1-\frac{v^2}{c^2}\left(\frac{cp}{\hbar\Omega l}\right)^2\sin^2\theta}
\label{GeneralRadiationRate}
\end{eqnarray}
where $v=\Omega R$  is the linear velocity of the circulating body; and $\theta$ is the polar angle of the emitted photon. The semiclassical approach is valid when $\kappa l\gg 1$. This is satisfied either for large angular momentum $l\gg 1$ or for large $\kappa$ which takes place when $v\sin\theta\ll c$.
For the radiation in plane ($\theta=\pi/2$)  satisfying the resonance condition $cp=\hbar \Omega l$ and for small velocity $v\ll c$ one obtains
\begin{equation}
w\sim \exp\left(-2l \ln\frac{c} {v}\right) \sim \left(\frac{\Omega R}{c}\right)^{2l}~,
\label{RateZS}
\end{equation}
which reproduces the Zeldovich result for the radiation by a rotating cylinder \cite{zeld}.

Equations (\ref{GeneralRadiationRate}) and (\ref{RateZS}) with $c$ substituted by speed of sound are valid for the radiation of phonons  by an external object circulating in the ultracold Bose condensates. The radiation occurs though the body is moving with the sub-critical velocity: the linear velocity of the object $\Omega R$ is smaller than the speed of sound.  In ultracold BEC, the role of a rotating object might be played by the moving optical potential made by the blue or red detuned laser  
\cite{Takeuchi2008}, or by a pair of like vortices which due to Magnus force rotate around their center of mass.

\end{document}